\documentclass[12pt]{iopart}

\usepackage{iopams}
\usepackage{graphicx}
\usepackage{dcolumn}
\usepackage{bm}
\usepackage{color}
\begin{document}
%\pagewiselinenumbers
\title[Static and dynamic polarizabilities of Yb-ion]{Accurate determination of black-body radiation shift, magic and tune-out wavelengths for the $\rm 6S_{1/2} \rightarrow 5D_{3/2}$ clock transition in Yb$^+$}

\author{A. Roy$^{1,2}$, S. De$^{1,2}$, Bindiya Arora$^{3}$, and B. K. Sahoo$^{4}$}
\address{$^1$CSIR-National Physical Laboratory, Dr. K. S.
Krishnan Marg, New Delhi-110012, India. \\
$^2$Academy of Scientific and Innovative Research, CSIR- National
Physical Laboratory Campus, New Delhi, India. \\
$^3$Department of physics, Guru Nanak Dev university, Amritsar,
Punjab-143005, India. \\
$^4$Atomic, Molecular and Optical Physics Division, Physical
Research Laboratory, Navrangpura, Ahmedabad-380009, India.}
\ead{royatish8@gmail.com} \vspace{10pt}
\begin{indented}
\item[]December 2016
\end{indented}

\begin{abstract}
We present precise values of the dipole polarizabilities ($\alpha$) of the ground $\rm [4f^{14}6s] ~ ^2S_{1/2}$ and metastable
$\rm [4f^{14} 5d] ~ ^2D_{3/2}$ states of Yb$^+$, that are 
%vital 
{\bf important} in reducing systematics in the clock frequency of the $\rm[4f^{14}6s] ~
^2S_{1/2} \rightarrow [4f^{14}5d] ~ ^2D_{3/2}$ transition. The static values of $\alpha$ for the ground and $\rm [4f^{14} 5d] ~ ^2D_{3/2}$
states are estimated to be $9.8(1) \times 10^{-40} \,\,\rm Jm^2V^{-2}$ and  $17.6(5) \times 10^{-40}\,\, \rm Jm^2V^{-2}$, respectively,
while the tensor contribution to the $\rm [4f^{14} 5d] ~ ^2D_{3/2}$ state as $- 12.3(3) \times 10^{-40}\,\, \rm Jm^2V^{-2}$ compared
to the experimental value $-13.6(2.2) \times 10^{-40}\,\,\rm Jm^2V^{-2}$. This corresponds to the differential scalar polarizability value
of the above transition as $-7.8$(5)$\,\times\, 10^{-40}\,\rm Jm^2 V^{-2}$ in contrast to the available experimental value
$-6.9$(1.4)$\,\times\, 10^{-40}$\,\, $\rm Jm^2V^{-2}$. This results in the black-body radiation (BBR) shift of the clock transition
as $-0.44(3)$ Hz at the room temperature, which is large as compared to the previously estimated values. Using the dynamic $\alpha$
values, we report the tune-out and magic wavelengths that could be of interest to subdue 
%major 
systematics due to the Stark shifts
and for constructing lattice optical clock using Yb$^+$.
\end{abstract}

\section{\label{sec:level0}Introduction}

\begin{figure}[t]
\begin{center}
\includegraphics[width=16.0cm, height=10.0cm]{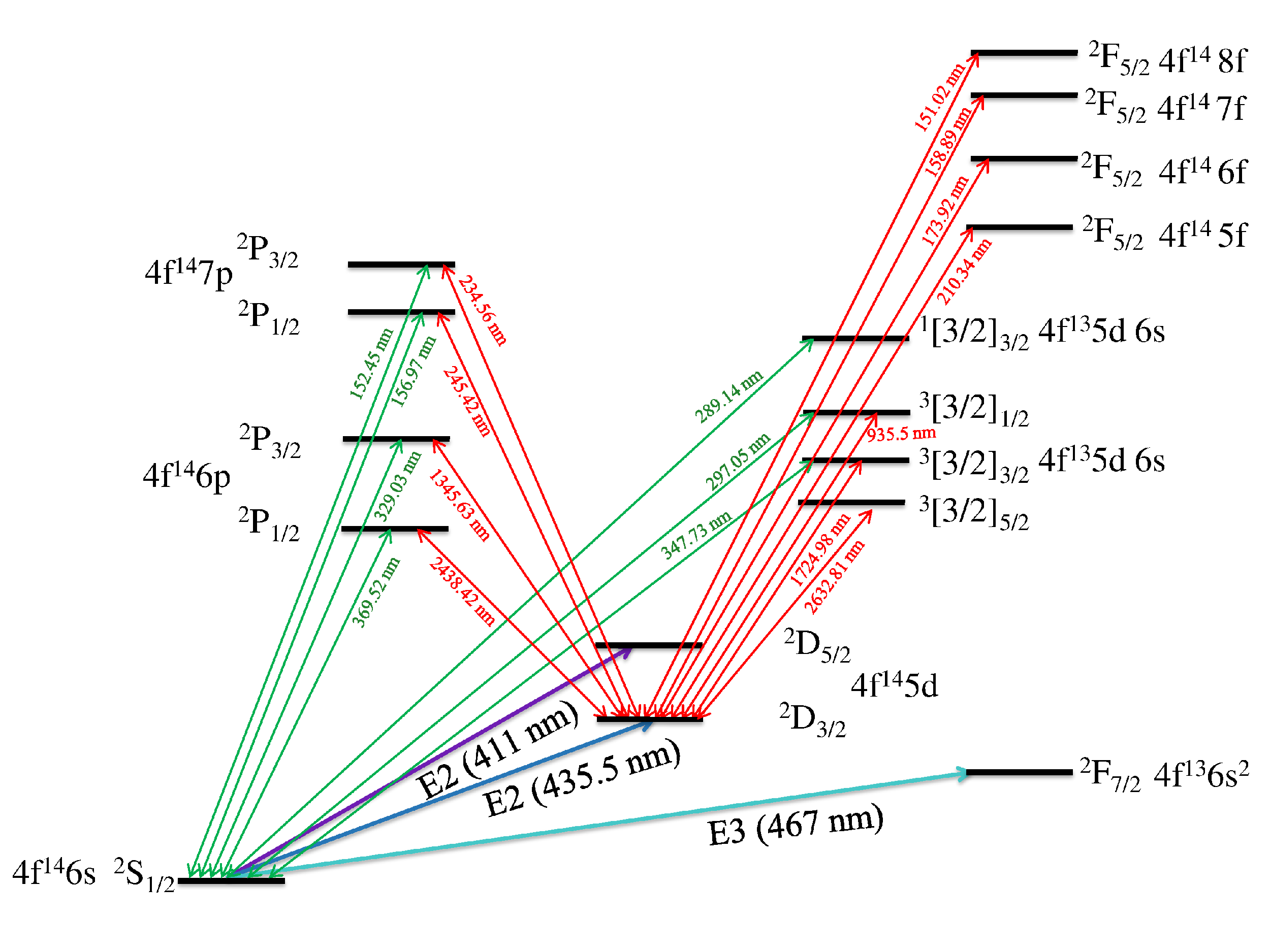}
\caption{Energy level diagram of $\rm Yb^+$: All the clock transitions and
low-lying states that are relevant for the experimental set-up and demonstrating
role of their importance in calculating dipole polarizabilities of the $\rm[4f^{14}6s] ~ ^2S_{1/2}$ and
$[4f^{14}5d] ~ ^2D_{3/2}$ states are highlighted here.
\label{eng_state}}
\end{center}
\end{figure}

Ytterbium-ion (Yb$^+$) is a unique system as its two quadrupole (E2) transitions, $\rm[4f^{14}6s] ~ ^2S_{1/2}\rightarrow
[4f^{14}5d] ~ ^2D_{3/2}$ and $\rm[4f^{14}6s] ~ ^2S_{1/2}\rightarrow [4f^{14}5d] ~ ^2D_{5/2}$, and one octupole
(E3) transition, $\rm [4f^{14}6s] ~  ^2S_{1/2}\rightarrow [4f^{13}6s^2] ~  ^2F_{7/2}$, as shown in Fig. \ref{eng_state} are
being undertaken for the optical frequency standards \cite{roberts,tamm,imai,atish,huntemann}. The $\rm[4f^{14}6s] ~
^2S_{1/2}\rightarrow [4f^{14}5d] ~ ^2D_{3/2}$ transition in Yb$^+$ has also been considered for probing parity nonconservation effect
\cite{bijaya}. Since the $\rm[4f^{14}6s] ~ ^2S_{1/2}\rightarrow [4f^{13}6s^2] ~ ^2F_{7/2}$ transition is strongly sensitive
to variation of the fine structure constant ($\alpha_e$) than the $\rm[4f^{14}6s] ~ ^2S_{1/2}\rightarrow [4f^{14}5d] ~ ^2D_{3/2}$
transition in Yb$^+$ \cite{dzuba1,dzuba2}, combining ratio of frequencies of these clock transitions with the clock frequencies of
Cs atom and combined frequencies of Hg$^+$ and Al$^+$ ions, limits on the variations of $\alpha_e$ and proton-to-electron mass ratio
are inferred \cite{godun}. In fact, all of the clock transitions in $\rm Yb^+$ seem to be sensitive to investigating possible
Lorentz symmetry violation \cite{dzubanat}. Owing to 
%overly 
long lifetime of the $\rm[4f^{13}6s^2] ~ ^2F_{7/2}$
state ($\geq$6 years), it makes the $\rm [4f^{14}6s] ~  ^2S_{1/2}\rightarrow [4f^{13}6s^2] ~  ^2F_{7/2}$ transition highly forbidden
and most suitable for the optical clock. On the other hand, lifetimes of the $\rm[4f^{14}5d] ~ ^2D_{3/2}$ and
$\rm[4f^{14}5d] ~ ^2D_{5/2}$ states are about 55 ms and 7 ms, respectively \cite{yu, nandy}. Hence, they are also suitable
for the optical clocks.

Optical atomic clocks are realized in two different ways. First,
in which the neutral atoms are localized in optical lattice and in
the other a singly charged ion is confined in a Paul trap. Optical
lattice clocks are capable of offering better statistics owing to
their large signal-to-noise ratio compared to a single trapped ion
clock. In contrast, using the sophisticated cooling and trapping
techniques, a trapped singly charged ion can be isolated from the
environmental perturbations which then acquires long coherence
time to the first order. Thereby, systematics of the measured clock frequencies can
be efficiently controlled and estimated in trapped ions, making
them suitable contenders for the optical atomic clocks \cite{neha_2016}. Precise
measurements of the differential black-body radiation (BBR) shifts, which are crucial factors in deciding the
uncertainties of the clock transitions, are extremely challenging. Analogously accurate estimates of these quantities demand for reliable
knowledge of the scalar components of the static polarizabilities.
Since precise values of the static scalar polarizabilities of the
atomic states associated with the $\rm[4f^{14}6s]~
^2S_{1/2}\rightarrow [4f^{14}5d] ~ ^2D_{3/2}$ clock transition in
Yb$^+$ are lacking, uncertainty in the BBR shift of this
transition is not known 
%meticulously
\cite{Polarizability_EFTF,tamm_eftf,experimental_polarization}.
Besides, consideration of ions in an optical lattice would be of interest for carrying out precise measurements on them as
has been suggested in~\cite{PLLiu_PRL_2015}. Few ions have already
been considered for experimental investigation either in an
optical dipole trap or in an optical lattice. For example, optical
dipole trapping of Mg$^+$  at the laser wavelength of 280 nm
loaded in a linear Paul trap was demonstrated for experimental
implication of the quantum simulations
\cite{ChSchnieder_Nature_2010}. The same group have also reported
an 1D optical lattice immersed in a 3D optically trapped Mg-ions
to explore their quantum dynamics \cite{MEnderlein_PRL_2012}.
Optical lattice of Yb$^+$ at the wavelength 377.2 nm has been
experimentally demonstrated for controlling their spin motion
couplings \cite{SDiang_PRL_2014,zhang_2017}. Adeptness of a trapped atomic
system is enhanced by using optical lattice operating at magic
wavelengths ($\lambda_{\rm magic}$), where the differential Stark
shift of a given transition is effectively nullified when the
lattice lasers are applied to the atomic systems. This was first
demonstrated by Katori {\it et. al.} for neutral Sr atoms
\cite{katori123}, following this a number of experimental and
theoretical works have been carried out in other neutral atoms
\cite{andrei,barber,bindiya1234}. However, measurement of
$\lambda_{\rm magic}$ at 395.7992(7) nm and 395.7990(7) nm in the
$^{40}$Ca$^+$ have been reported recently \cite{PLLiu_PRL_2015}.
After that these values have been investigated theoretically in
other alkaline-earth ions \cite{JKaur_PRA_2015}. Similarly, the
Stark shift of an energy level is zero at the tune-out wavelengths
($\rm\lambda_T$) so the atomic systems can be free from the 
%dominant
systematics due to the Stark shifts if they are trapped using
lasers at these values. Measurements of the tune-out wavelengths may be used
as high-precision benchmark tests of theory \cite{tuneout}. Moreover, it is possible to combine measurements of magic wavelengths and lightshifts with high accuracy theoretical calculations to infer the values of some of the E1 matrix elements which are difficult to measure
otherwise~\cite{PLLiu_PRL_2015,sherman,bkslight}.

In this article, we present accurate values of the static and dynamic dipole polarizabilities of the $\rm[4f^{14}6s]~ ^2 S_{1/2}$
and $\rm[4f^{14}5d] ~ ^2D_{3/2}$ states in $^{171}$Yb$^+$. For their evaluations, we have used 
%precisely 
known electric dipole (E1)
matrix elements extracted from the measurements of the lifetimes of some of the low-lying atomic states calculated elsewhere using the
relativistic coupled-cluster (RCC) method. We have estimated core correlation effects and other contributions employing other variants
of relativistic many-body methods. Using these polarizabilities, we determine the BBR shift, tune-out and magic wavelengths of the
$\rm[4f^{14}6s] ~ ^2S_{1/2}-\rm[4f^{14}5d] ~ ^2 D_{3/2}$ clock transition in $^{171}$Yb$^+$. The BBR shift of this clock
transition is calculated as $-0.44(3)$ Hz at the 300 K. We also found several magic and tune-out wavelengths, among
which suitable choices can be made for performing different high-precision experiments depending on their applications and
availability of the lasers with the required power.

\section{\label{sec:level1}Theory}
\textbf{The Stark shift in the energy level of the $n^{th}$ hyperfine level with angular momentum $F_n$
and its magnetic component $M_{F_n}$ placed in an uniform oscillating electric field of a laser beam
$\mbox{\boldmath${\cal E}$}(t)= \frac{1}{2}{\cal E} \hat{\mbox{\boldmath$\varepsilon$}} e^{-(\iota\omega t+-\textbf{k}.\textbf{r})}+c.c.$, with ${\cal E}$ 
being the amplitude, $\hat{\mbox{\boldmath$\varepsilon$}}$ is the polarization vector of the electric field and $c.c.$ referring to the complex conjugate of the former term, 
oscillating at frequency $\omega$ is given by }
\begin{eqnarray}
\Delta E^{\rm Stark}_{F_n} \approx
-\frac{1}{4}\alpha_{F_n}(\omega){\cal E}^2.\label{basic} ,
\end{eqnarray}
where $\alpha_{F_n}(\omega)$ is the frequency dependent or dynamic polarizability of the state and given by
\begin{eqnarray}
\alpha_{F_n}(\omega) &=& \alpha_{F_n}^{(0)} (\omega) +
\alpha_{F_n}^{(1)} (\omega ) \frac{A \cos\theta_k M_{F_n}}{2F_n}
+ \alpha_{F_n}^{(2)} (\omega ) \nonumber \\
& \times &  \left( \frac{3\cos^2\theta_p-1}{2} \right)
 \left [ \frac{3M_{F_n}^2-F_n(F_n+1)}{F_n(2F_n-1)} \right ]. \label{total}
\end{eqnarray}
Here $\alpha_{F_n}^{(i)}(\omega)$ are the scalar,
vector and tensor components of the frequency dependent
polarizability for i = 0, 1, and 2 respectively, $A$ represents degree of
polarization, $\theta_k$ is the angle between quantization axis
and wave vector, and $\theta_p$ is the angle between quantization
axis and direction of polarization. In the presence
of magnetic field, $\cos\theta_k$ and $\cos^2\theta_p$ can have
any values depending on the direction of applied magnetic field.
In the absence of magnetic field, $\cos\theta_k = 0$  and
$\cos^2\theta_p = 1$  for the linearly polarized light, where
polarization vector is assumed to be along the quantization axis.
However it yields $\cos\theta_k = 1$ and $\cos^2\theta_p = 0$ for
the circularly polarized light, where the wave vector is assumed to be
along the quantization axis. Here we present results for both
polarizations of the light assuming absence of the magnetic field. The component of
the polarizability can be expressed as ~\cite{beloy_2009, Dzuba_2010}
\begin{eqnarray}
\alpha_{F_n}^{(0)}(\omega)&=&\alpha^{(0)}_{J_n}(\omega), \\
\alpha_{F_n}^{(1)}(\omega)&=&(-1)^{J_n+F_n+I+1}  \left\{
\begin{array}{ccc}
                             F_n & J_n & I\\
                          J_n & F_n &1
                         \end{array}\right\}   \nonumber \\ &\times & \sqrt{\frac{F_n(2F_n+1)(2J_n+1)(J_n+1)}{J_n(F_n+1)}}
              \alpha^{(1)}_{J_n}(\omega) \ \
\end{eqnarray}
and
\begin{eqnarray}
\alpha_{F_n}^{(2)}(\omega)&=&(-1)^{J_n+F_n+I}  \left\{
\begin{array}{ccc}
                                            F_n& J_n & I\\
                                            J_n & F_n &2
                                           \end{array}\right\}  \nonumber \\
                                           & \times & \sqrt{\frac{F_n(2F_n-1)(2F_n+1)}{(2F_n+3)(F_n+1)}} \nonumber \\
& \times & \sqrt{\frac{(2J_n+3)(2J_n+1)(J_n+1)}{J_n(2J_n-1)}}
\alpha^{(2)}_{J_n}(\omega) .
\end{eqnarray}
Here, $J_n$ is the total angular momentum of the atomic state, $I$
is the nuclear spin and $\alpha^{(i)}_{J_n}$ for i = 0, 1, 2 are the
scalar, vector and tensor components of the atomic dipole polarizability
which are given by~\cite{Manakov_1986}
\begin{eqnarray}
\alpha^{(0)}_{J_n}(\omega)&=&\frac{2}{3(2J_n+1)} \sum_{k \ne n} \frac{(E_n -E_k)|\langle J_n||{\bf D}||J_k \rangle|^2} {\omega^2 - (E_n -E_k)^2}, \label{scalar} \\
\alpha^{(1)}_{J_n}(\omega)&=& \sqrt{\frac{24J_n}{(J_n+1)(2J_n+1)}}\sum_{k \ne n}(-1)^{J_n+J_k}  \nonumber \\
              & & \times \left\{ \begin{array}{ccc}
                             J_n& 1 & J_n\\
                          1 & J_k &1
                         \end{array}\right\} \frac{ \omega |\langle J_n||{\bf D}|| J_k \rangle|^2}  {\omega^2 - (E_n -E_k)^2 }  \label{vector}
\end{eqnarray}
and
\begin{eqnarray}
\alpha^{(2)}_{J_n}(\omega) &=& \sqrt{\frac{40
J_n(2J_n-1)}{3(J_n+1)(2J_n+3)(2J_n+1)}}
 \sum_{k \ne n}(-1)^{J_n+J_k} \nonumber \\ &\times &
                                  \left\{ \begin{array}{ccc}
                                            J_n& 2 & J_n\\
                                            1 & J_k &1
                                           \end{array}\right\} \frac{ (E_n -E_k)|\langle J_n||{\bf D}||J_k \rangle|^2} {\omega^2 - (E_n -E_k)^2}, \label{tensor}
\end{eqnarray}
where $J_m$ and $E_m$ are the total angular momentum and energy,
respectively, for $m = n, \, k$ states and $\langle J_n||{\bf
D}||J_k\rangle$ are the reduced E1 matrix elements.

The BBR shift experienced by a state resulting from radiations with all possible frequencies at a temperature $T$ (in Kelvin (K))
relative to the room temperature can be estimated using the expression \cite{porsev_2006}
\begin{eqnarray}
\Delta E_F^{BBR} \simeq -\frac{1}{2} (831.9\,\,\rm{V/m})^2
\left[\frac{{\it{T}(\rm K)}}{300}\right]^4 \alpha_F^{(0)}(0){\bf(1+ \eta)} ,
\end{eqnarray}
 where $\eta$ is the dynamic fractional correction factor and is defined as
\begin{eqnarray}
 \eta = \frac{(80/63)\pi^2 T^2}{(J_n+1)\alpha_F^{(0)}(0)} \sum_k \frac{ |\langle J_n||{\bf D}|| J_k \rangle|^2}  {(E_n -E_k)^3 }
 \left (1+ \frac{21 \pi^2 T^2}{5(E_n -E_k)^2 }+ \frac{336 \pi^4 T^4}{11(E_n -E_k)^4 }\right ). \ \ \ \ \ \
 \label{dync}
\end{eqnarray}

The differential BBR shift of a transition is, thus, the difference of the BBR shifts of the states associated with that transition. We
use atomic units (a.u.) throughout the article unless stated otherwise. The  
$\alpha_{F_n}$ values can be converted from a.u.
to $\rm {Hz ~m^2 V^{-2}}$ and $\rm {J m^2 V^{-2}}$ by dividing with factors $2.48832 \times 10^{-8}$ and  6.064\,$\times$\, $10^{40}$,
respectively.

The differential AC-Stark shift for a transition between states
with hyperfine moments $F$ and $F'$ is given by
\begin{eqnarray}
\delta (\Delta E^{Stark}_{FF'}) &\approx&\Delta E_F^{Stark}-\Delta E^{Stark}_{F'} \nonumber \\
%                      &=&-\frac{1}{2}\left[\alpha_{F}(\omega)-\alpha_{F'}(\omega)\right]{\cal E}^2(\omega) \nonumber \\
                      &=&-\frac{1}{2} \delta \alpha_{FF'}(\omega) {\cal E}^2(\omega). \label{diff_polar}
\end{eqnarray}
In order to find out $\rm\lambda_{magic}$ of a transition, it is
imperative to identify the corresponding values of $\omega$ at
which the differential Stark shift $\delta (\Delta
E^{Stark}_{FF'})$ is zero for any finite electric field strength
${\cal E}$. By suitably choosing polarization of the electric
field and azimuthal sublevels, $\rm\lambda_{magic}$ can be
determined for different hyperfine levels of a transition.
In fact, we had demonstrated in a recent work that by suitably deciding a trap geometry the Stark shift of an atomic state or
differential Stark shift of a transition can be obtained which can be independent of the vector and tensor components of the states
involved \cite{sukhjit}. Assuming such a trapping configuration can be achieved for trapping the Yb$^{+}$ ion, we
also give $\rm\lambda_{magic}$ for the aforementioned clock transition. This {\bf comprehensive trapping scheme} could be useful for avoiding hyperfine level selective
trapping of $^{171}$Yb$^+$. We also evaluate $\rm\lambda_T$ values for which $\alpha_F(\omega)$ can become zero independently for the
$\rm [4f^{14}6s]\,\,^2S_{1/2}$ and  $\rm [4f^{14}5d]\,\,^2D_{3/2}$ states.

\begin{table}
\centering {\caption{ Contributions to the static values of
$\alpha_{J_n}^{(i=0,2)}$ for the $\rm{[4f^{14}6s]\,\,^2S_{1/2}}$
and $\rm [4f^{14}5d] ~ ^2D_{3/2}$ states in Yb$^{+}$ are given in a.u. Absolute values of the dominantly
contributing reduced E1 matrix elements (in a.u.) and resonance
wavelengths in vacuum ($\lambda_{\rm{res}}$ in nm) are also
listed. Values quoted in bold fonts are taken from other work as mentioned below.
% Refs.\cite{[4],[2],[3],spectrochimica_2010}.
%against the CCSD results given in the parentheses just below the respective matrix element.
}
 \label{polar_ground}
%\begin{ruledtabular}
\begin{tabular}{lllrr}
\hline
\hline
 Contribution     & \,\,\,$\lambda_{\rm{res}}$~\cite{Nist_data} & \,\,\,\,\,\,\,E1             & $ \alpha^{(0)}_{J_n}$\,\,  & $ \alpha^{(2)}_{J_n}$\,\,  \\
 \hline
 & & & &  \\
 $ \rm[4f^{14} 6p] ~  ^2 P_{1/2} $    &  369.52 ~ \,             & {\bf2.471(3)} \cite{[2],[3]}    & 16.51(4)                            \\
 $\rm[4f^{13}5d6s] ~ ^3[3/2]_{3/2}$    &     347.73~ \,    & {\bf1.10} \cite{[4]} ~  ~ \,          & {3.1}       &       \\
 $ \rm[4f^{14}6p]  ~ ^2 P_{3/2} $      &   329.03 ~ \,      & {\bf3.36(3)} \cite{[2],[3]} & 27.2(5)                            \\
$\rm[4f^{13}5d6s] ~ ^3[3/2]_{1/2}$  &  297.05 ~ \,       &  {\bf0.82(2)} \cite{spectrochimica_2010,J_phy_biemont_1998} ~  ~ \,         &  {1.46(3)}     &      \\
  $\rm[4f^{13}5d6s] ~ ^1[3/2]_{3/2}$ &  289.14 ~ \,       &  {\bf1.27(2)} \cite{spectrochimica_2010}   ~  ~ \,      &  {3.45(3)}      &       \\
 $ \rm[4f^{14}7p]  ~ ^2 P_{1/2}$       &  156.97 ~ \,      & 0.08(1) ~  ~ \,    & 0.0008(1)                            \\
 $ \rm[4f^{14}7p]  ~ ^2 P_{3/2}$        &   152.45 ~ \,     & 0.11(1) ~  ~ \,     & 0.014(2)                            \\
  $\alpha^{(i,c)}$        &  &              & 7.7(7)  &                                  \\
 $\alpha^{(i,cv)}_{6s^2S_{1/2}}$        &  &              & $-0.16(2)$  &                                    \\
 ${\rm{Tail}}(\alpha^{(i,v)}_{6s\,\,^2S_{1/2}})$      &  &         & 0.046(15)     &                                 \\
Total                              &   &           & 59.3(8)   &                                  \\
 Ref. \cite{MBPT}  &&&62.04\\
 Ref. \cite{Polarizability_EFTF}                         &   &              &  47(9)$^a$    &   \\
 & & \\
  $\rm[4f^{13}5d6s] ~ ^3[3/2]_{5/2}$   & 2632.81  & {\bf0.00075} \cite{spectrochimica_2010}  & 0.0                                  & 0.0   \\
  $\rm[4f^{14}6p] ~  ^2 P_{1/2} $  &  2438.42   & {\bf2.97(4)} \cite{[2],[3]} & 79(2)                              & $-79(2)$ \\
$\rm[4f^{13}5d6s] ~ ^3[3/2]_{3/2}$  & 1724.98  & {\bf0.27} \cite{[4]}     & 0.46                                & 0.37 \\
 $\rm[4f^{14}6p] ~  ^2 P_{3/2} $  &   1345.63      & 1.31(2)   & 8.45(26)                           & 6.76(21) \\
 $\rm[4f^{13}5d6s] ~ ^3[3/2]_{1/2}$ & 935.19        & {\bf0.62(1)}\cite{spectrochimica_2010,J_phy_biemont_1998}  & 1.31(4)                            & -1.31(4) \\
 $\rm[4f^{14}7p]  ~ ^2 P_{1/2}$    &    245.42      & 0.14(2)   & 0.018(5)                           & $-0.018(5)$ \\
 $\rm[4f^{14}7p]  ~ ^2 P_{3/2}$    &    234.56      & 0.014(5)   & 0.0002(1)                          & 0.0001(1)  \\
  $\rm[4f^{14} 5f]  ~ ^2 F_{5/2}$   &    210.34      & 2.43(4)     & 4.54(15)                           & $-0.91(3)$   \\
   $\rm[4f^{14}6f]  ~ ^2 F_{5/2}$  &    173.92      & 1.47(2)  & 1.38(4)                             & $-0.275(7)$  \\
    $\rm[4f^{14}7f]  ~ ^2 F_{5/2}$   &   158.89   & 0.93(1)    & 0.503(11)                           & $-0.101(2)$   \\
     $\rm[4f^{14}8f]  ~ ^2 F_{5/2}$ &   151.02      & 0.35(1)  & 0.068(4)                            & $-0.014(1)$  \\
  $\alpha^{(i,c)}$  & &                & 7.7(7)                                 & 0.0      \\
 $\alpha^{(i,cv)}_{5d^2D_{3/2}}$ &  &                & $-0.43(3)$                             & $-0.08(1)$   \\
 ${\rm{Tail}}(\alpha^{(i,v)}_{5d\,\,^2D_{3/2}})$ &&                & 4(2)                                 & $-0.83(25)$  \\
 Total                         & &              & 107(3)                                 & $-75(2)$     \\
 Ref. \cite{Polarizability_EFTF}                     &  &            & 90(17)$^a$                                 &    \\
  Ref. \cite{experimental_polarization}                     &  &            &                                  & $-82(13)$   \\
\hline \hline
 \end{tabular}}
%\end{ruledtabular}
\noindent{\,\,\,\,\,\,\,\,\,\,\,\,\,\,\,\,\,\,\,\,\,\,\,\,\,\,\,\,\,\,\,\,\,\,\,\,\,$^a$Note: Estimated by combining measured differential
scalar polarizability $\alpha^{(0)}_{6s \,\,\,^2S_{1/2}}-\alpha^{(0)}_{5d \,\,\,^2D_{3/2}}$ from
Ref.~\cite{experimental_polarization} and E1 matrix elements from the measured lifetimes of many states as listed in Ref.~\cite{Polarizability_EFTF}.}
\end{table}
%%%%%%%%%%%%%%%%%%%%%%%%%%%%%%%%%%%%%%%%%%%%%%%%%%%%%%%%%%%%%%%%%%%%%%%%%%%%%%%%%%%%%%%%%%%%%%%%%%%%%

%%%%%%%%%%%%%%%%%%%%%%%%%%%%%%%%%%%%%%%%%%%%%%%%%%%%%%%%%%%%%%%%%%%%%%%%%%%%%%%%%%%%%%%%%%%%%%%%%%%%%

\section{\label{sec:level2}Method of calculations }

We divide contributions to each $\alpha^{(i)}_{J_n}$ component
broadly into three parts, as described in Refs.
\cite{JKaur_PRA_2015,b.arora.nandy}, which is expressed as
\begin{eqnarray}
\alpha^{(i)}_{J_n} = \alpha^{(i,c)} + \alpha^{(i,cv)}_{J_n} +
\alpha^{(i,v)}_{J_n}\,\,\,\,. \label{correlation}
\end{eqnarray}
 The superscripts $c$, $cv$ and $v$ in the
parentheses are known as core (independent of $J_n$), core-valence
and valence contributions, respectively
\cite{JKaur_PRA_2015,b.arora.nandy}. When the core of the electron
distribution has an inert gas configuration, core and core-valence
contributions usually come out to be much smaller than the valence
contribution
\cite{bindiya1234,JKaur_PRA_2015,b.arora.nandy,MBPT}. This has
also been observed earlier in the ground state polarizability study of Yb$^+$
\cite{MBPT}. Again, it is possible to use the reduced E1 matrix
elements among the low-lying bound states directly in Eqs.
(\ref{scalar}-\ref{tensor}) in a sum-over-states approach to
estimate the dominant contribution to $\alpha^{(i,v)}_{J_n}$. We
refer to this as ``Main'' valence correlation contribution and the
rest that comes from the other higher level excited  states as ``Tail''
contribution to $\alpha^{(i,v)}_{J_n}$.

For accurate determination of the E1 matrix elements between many low-lying states, we employ the relativistic coupled-cluster (RCC) method
by expressing the wave function ($\vert \Psi_v \rangle$) of an atomic state that has the closed core
$\rm[5p^6]$ and a valence orbital $v$ as
\begin{eqnarray}
 \vert \Psi_v \rangle = e^T \{ 1+S_v \} \vert \Phi_v \rangle
 \label{cceq}
\end{eqnarray}
where $\vert \Phi_v \rangle$ is a reference state, which is
defined as $\vert \Phi_v \rangle = a_v^{\dagger} \vert \Phi_0
\rangle$ with the Dirac-Hartree-Fock (DHF) wave function $\vert \Phi_0 \rangle$
of the closed-core, $T$ and $S_v$ are the RCC excitation operators
that excite electrons from the core and core along with the
valence orbital respectively. In this work, we have only accounted
for the single and double excitations which are denoted using the
subscripts $1$ and $2$ respectively in the RCC operators as
\begin{eqnarray}
 T=T_1 +T_2 \ \ \ {\rm and} \ \ \ S_v = S_{1v} + S_{2v}.
\end{eqnarray}
These are known as coupled-cluster singles and doubles (CCSD)
method in the literature. Amplitudes of these operators are
evaluated using the equations
\begin{eqnarray}
 \langle \Phi_0^* \vert \overline{H}  \vert \Phi_0 \rangle &=& 0
\label{eqt}
 \end{eqnarray}
and
\begin{eqnarray}
 \langle \Phi_v^* \vert \big ( \overline{H} - E_v \big ) S_v \vert \Phi_v \rangle &=&  - \langle \Phi_v^* \vert \overline{H} \vert \Phi_v \rangle ,
\label{eqsv}
 \end{eqnarray}
where $\vert \Phi_0^* \rangle$ and $\vert \Phi_v^* \rangle$ are
the excited state configurations, here up to doubles, with respect
to the DHF states $\vert \Phi_0 \rangle$ and $\vert \Phi_v
\rangle$, respectively, and $\overline{H}= \big ( H e^T \big )_l$
with subscript $l$ representing the linked terms only. $E_v$
is the energy eigenvalue of the $\vert \Psi_v \rangle$ state and
is determined by using the expression
\begin{eqnarray}
 E_v  = \langle \Phi_v \vert \overline{H} \left \{ 1+S_v \right \} \vert \Phi_v \rangle .
 \label{eqeng}
\end{eqnarray}
Both Eqns. (\ref{eqsv}) and (\ref{eqeng}) need  to be solved simultaneously for obtaining solutions of interdependent variables
in a self-consistent approach. At the same level of approximation in the excitations, a truncated RCC method can incorporate more
contributions as compared to a truncated configuration-interaction (CI) method. For example, the lowest order mixing of
$[4f^{13}5d6s]\,\,^3[3/2]$ with $J=1/2$ configuration and the DHF wave function of the $[4f^{14}6p]\,\,^2P_{1/2}$ configuration is
carried out by the $S_{2v}$ operator while calculating the $[4f^{14}6p]\,\,^2P_{1/2}$ atomic state function. Similar mixing also
takes place among the $J=3/2$ configurations. Moreover, $\overline{H}$ in the CCSD method contains many non-linear terms that can take into
account contributions from the higher level excitations, such as triples and quadruples, at the same level of excitation approximation
of the CI method. As a matter of fact, the above mentioned configuration mixing are incorporated through the non-linear terms of the
CCSD method and they take care of higher order correlation contributions. This is a unique feature of a truncated RCC method in
contrast to other truncated many-body methods owing to the exponential form of expression in Eq. (\ref{cceq}). In fact, it has also
been employed to calculate many atomic properties of Yb$^+$ accurately in the past \cite{bijaya,nandy}.

After obtaining amplitudes using the above equations, the
transition matrix element of the dipole operator $D$ between the
states $\vert \Psi_i \rangle$ and $\vert \Psi_f \rangle$ is
evaluated using the expression {\small
\begin{eqnarray}
\frac{\langle \Psi_f \vert D \vert \Psi_i \rangle} {\sqrt{\langle
\Psi_f \vert \Psi_f\rangle  \langle \Psi_i|\Psi_i\rangle}}
      &=& \frac{\langle\Phi_f|\tilde{D}_{fi}|\Phi_i\rangle}{\sqrt{\langle\Phi_f|\{1+\tilde{N}_f\}|\Phi_f\rangle\langle\Phi_i|\{1+\tilde{N}_i\}|\Phi_i\rangle}} , \nonumber \\
\end{eqnarray}}
where $\tilde{D}_{fi}=\{1+S_f^{\dagger} \} e^{T^{\dagger}} D e^T
\{1+S_{i}\}$ and $\tilde{N}_{k=f,i}=\{1+S_k^{\dagger} \}
e^{T^{\dagger}} e^T \{1+S_{k}\}$. The above expression involves two
non-terminating series in the numerator and denominator, which are
$e^{T^{\dagger}} D e^T$ and $e^{T^{\dagger}} e^T$ respectively,
and calculation of these terms are described in our previous works
\cite{nandy}. The ``Main($\alpha_{J_n}^{(i,v)}$)''
contributions have been calculated by combining the E1 matrix elements obtained using
the above expression with the experimental energies as listed in
Ref. \cite{Nist_data}.

Contributions from higher excited states including continuum to
$\alpha_{J_n}^{(i,v)}$, denoted as
``Tail($\alpha_{J_n}^{(i,v)}$)'', are estimated approximately in
the DHF method using the expression
\begin{eqnarray}
\alpha_{J_n}^{(i,v)} ( \omega) &=& \frac{2}{3(2J_n+1)} \sum_{K >
n} \frac{ (\epsilon_K-\epsilon_n) |\langle
J_n || {\bf D} || J_K \rangle_{DHF} |^2} {(\epsilon_n - \epsilon_K)^2 - \omega^2}, \nonumber \\
\end{eqnarray}
where $\langle J_n || {\bf D} || J_K \rangle_{DHF}$ are obtained
using the DHF wave functions, $K>n$ corresponds to the excited
states including continuum whose matrix elements are not accounted
in the Main($\alpha_{J_n}^{(i,v)}$) contribution, and
$\epsilon_{n,K}$ refers to the DHF energies.

Similarly, the core-valence contributions $\alpha_{J_n}^{(i,cv)}$
are also obtained through the DHF method approximation using the expression
\begin{eqnarray}
\alpha_{J_n}^{(i,cv)} ( \omega) &=& \frac{2}{3(2J_n+1)}
\sum_{K}^{N_c} \frac{ (\epsilon_K-\epsilon_n) |\langle
J_n || {\bf D} || J_K \rangle_{DHF} |^2} {(\epsilon_n - \epsilon_K)^2 - \omega^2}, \nonumber \\
\end{eqnarray}
where $N_c$ is the total number of electrons in the closed-core of
the Yb$^+$. The static scalar core contributions
$\alpha^{(0,c)} (0)$ in Yb$^+$, which is equal to the ground state
polarizability of Yb$^{2+}$, are evaluated using the DHF,
random-phase approximation (RPA) and a perturbed CCSD method considering a dipole operator $D$ as an external
perturbation as described in Ref. \cite{yashpal}. The trends in these results demonstrate role of electron correlation
effects for its accurate evaluation. The DHF and RPA contributions are subsumed within the CCSD result by the formulation
of the theory and are considered as the most precise calculation. Comparison between the DHF and CCSD results than between the
RPA and CCSD show small differences between these results while the CCSD method is computationally very expensive. As a matter of fact,
it is not possible to apply the CCSD method to determine core contribution to the dynamic polarizabilities for the estimates of the
$\rm\lambda_{magic}$ and $\rm\lambda_T$ values. Therefore, we employ the DHF method to calculate the core contributions to the dynamic
polarizabilities. Nevertheless, core contributions cancel out completely in the determination of $\rm\lambda_{magic}$. Hence, accuracies
in those values are not affected by the use of core correlations from the DHF method while the $\rm\lambda_T$ values can be estimated
within reasonable accuracy.

%%%%%%%%%%%%%%%%%%%%%%%%%%%%%%%%%%%%%%%%%%%%%%%%%%%%%%%%%%%%%%%%%%%%%%%%%%%%%%%%%%%
\begin{table*}
\centering {\caption{The $\lambda_{\rm{magic}}$ values in nm and their corresponding
$\alpha_{F_n}$ values in a.u. values among different $M_{F_n}$ sub-levels of the
${\rm[4f^{14}6s] ~ ^2S_{1/2}}(F = 0) \rightarrow {\rm
[4f^{14}5d] ~ ^2D_{3/2}}(F = 2)$ clock transition in
$^{171}$Yb$^+$ are listed for both the linearly and right-circularly
polarized light. \label{magic} }
% \begin{ruledtabular}
\begin{tabular}{lrrr|lrrr}
\hline
\hline
\multicolumn{4}{c|}{Linear} &\multicolumn{4}{c}{Circular} \\
\hline
 & ${M_F}$ & $\lambda_{\rm{magic}}$\,\,\,   &$\alpha_{F_n}$\,\,\,  & & ${M_F}$ & $\lambda_{\rm{magic}}$\,\,\,  &$\alpha_{F_n}$\,\,\,   \\
 \hline
   & 0 \,\,\,\,       & 1725.4(1)   & 62(2) & & -2\,\,\,\,       & 1726.5(1)   & 60(5)\\
        &          &  1356.3(3)   & 63(2)  &&          & 1381.8(6)    & 62(2) \\
        &          & 951.8(5)     & 66(2)  &&          & 948.9(2)           & 64(2)        \\
        &          &  357.3(1)  & 21.03(1) &   &          & 358.1(2)   & -6.2(3) \\
         &          &  344.9(1)  & 21.81(4) & &          & 345.2(2)   & -5.1(3) \\
           &       &  298.8(3)  & 25.68(3) &  &          & 299.1(1)   & -1.31(6) \\
         &          &  291.8(1)  & 26.52(6) &  &          & 292.2(1)   & -0.80(1) \\
         &          &  245.4(1)  & -53.43(3) &  &          & 243.5(1)   & -55.3(2)\\
        & $\pm1$\,\,\,\,   & 1726.6(1)   & 61(2) & &-1\,\,\,\,& 1728.8(3)     & 61(2) \\
        &          & 1382(1)    & 63(2) & &          & 1418(2)    & 62(2)   \\
        &          & 948(1)     & 65(2) & &          & 944.4(2)           & 67(3)           \\
        &          & 357.3(2)   & 20.61(3) & &          & 357.8(1)    & 5.7(1) \\
         &          &  344.9(1)  & 21.01(2)& &          & 345.1(1)    & 6.4(2)  \\
        &          &  298.8(1)  & 24.62(6) & &          & 299.0(1)    & 9.41(5) \\
        &          &  291.8(1)  & 25.43(5) &  &          & 292.1(1)    & 9.89(7)\\
        &          &  245.4(1)  & -53.59(6) &     & 0\,\,\,\,        & 1735.4(2)     & 61.4(7) \\
        &$\pm2$\,\,\,\,    & 1788(6)    & 63(1) &    &          & 1479(5)    & 62(5)\\
        &                   & 1619(5)   & 59(1)        &    &         & 939.7(2) & 67(2) \\
        &                  & 357.4(2)    & 18.0(2)  &    &          & 357.4(2)  & 18.8(2)  \\
        &                  & 345.1(2)     & 18.42(2)  &               & &  345.0(2)    & 19.3(3) \\
         &                 & 298.8(1)     & 20.63(5)  & &          & 298.8(1)    & 21.96(5)\\
          &                & 291.9(1)    & 21.20(5)   &  &          & 291.9(1)    & 22.50(7) \\
                &          &             &             &  &       1\,\,\,\,   & 1657(15)    & 61(2)\\
&&&&        &         & 356.9(2)    & 32.1(6) \\
&&&&        &                  & 344.8(2)      & 32.5(3) \\
&&&&        &                  & 298.7(1)       & 35.30(5) \\
&&&&        &                  & 291.7(1)       & 35.84(5)  \\
&&&&        & 2\,\,\,\,        & 1204(45)     &  64(2)    \\
&&&&        &                  & 356.5(2)      & 46.6(9)  \\
&&&&        &                  & 344.7(2)      & 47.0(5)   \\
&&&&        &                  & 298.5(1)      & 50.31(7)  \\
&&&&        &                  & 291.6(1)    &   51.17(9) \\
\hline \hline
\end{tabular}}
% \end{ruledtabular}
\end{table*}

%%%%%%%%%%%%%%%%%%%%%%%%%%%%%%%%%%%%%%%%%%%%%%%%%%%%%%%%%%%%%%%%%%%%%%%%%%%%%%%%%%%%%%
\begin{figure*}[t]
\begin{center}
\includegraphics[width=16cm,height=12.0cm]{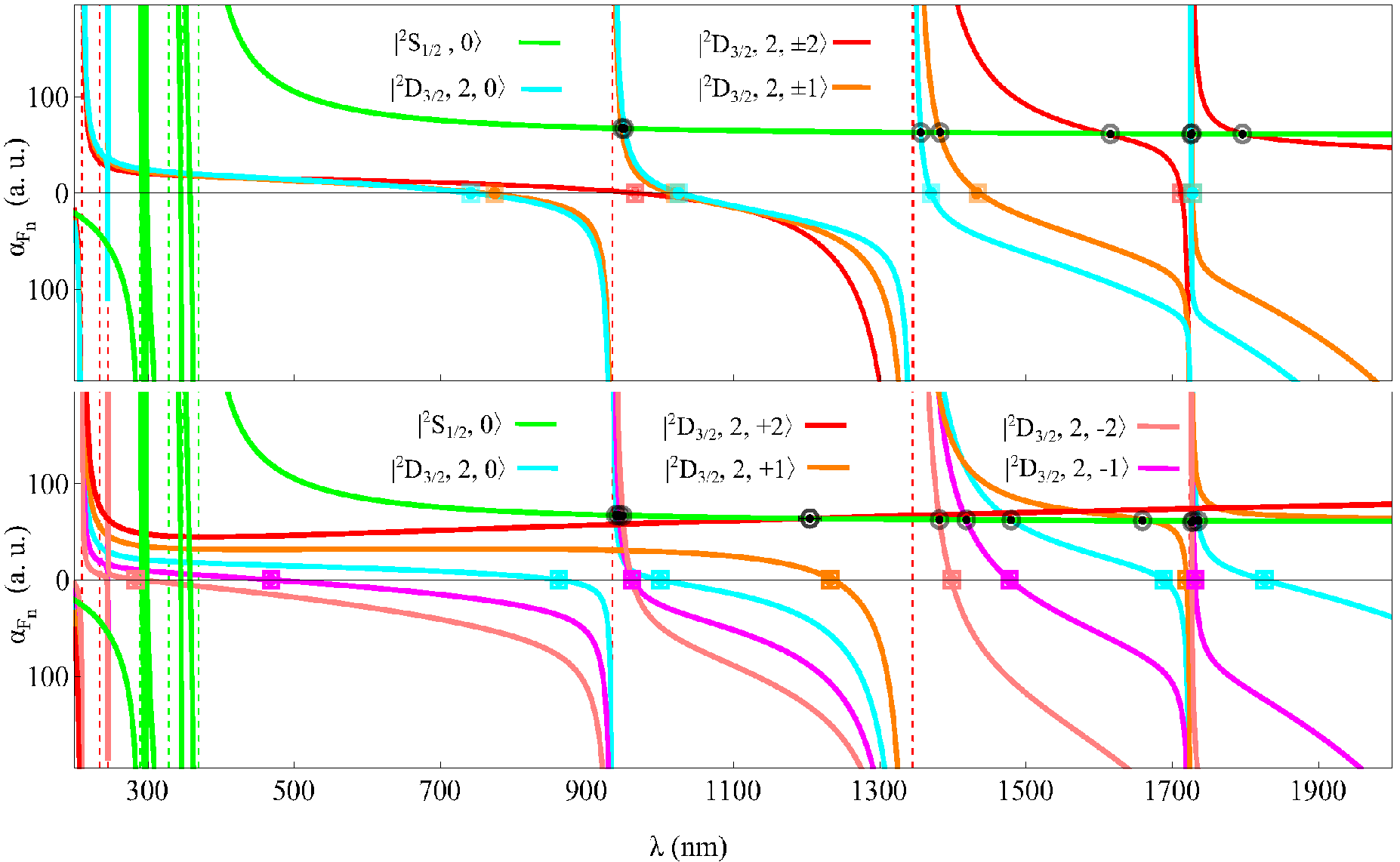}
\end{center}
\caption{Wavelength dependent dynamic dipole polarizabilities of
the ${\rm[4f^{14}6s] ~ ^2S_{1/2}}(F = 0)$ state (denoted
by $|\rm ^2S_{1/2}, 0 \rangle$) and for all possible $M_{F}$
sublevels of the ${\rm [4f^{14}5d] ~ ^2D_{3/2}}(F = 2)$
state (denoted by $|{\rm ^2D_{3/2}}, 2, \rm M \rangle$) in
$^{171}$Yb$^+$ using (a) linearly and (b) right-circularly
polarized light. The $\lambda_{\rm{magic}}$ and $\rm\lambda_T$ values
are indicated in dots which are encircled by circles and squares, respectively. The vertical dashed lines signify the resonances from the ground and ${\rm [4f^{14}5d] ~ ^2D_{3/2}}(F = 2)$  states in green and red colors, respectively.
\label{linear1}}
\end{figure*}

\section{\label{sec:level3}Results and Discussion}

The obtained static polarizabilities for the $\rm[4f^{14} 6s]\,\,^2S_{1/2}$ and $\rm[4f^{14} 5d]\,\,^2D_{3/2}$ states and estimated BBR
shifts associated with these states are presented in this section. It can be understood from Fig.~\ref{eng_state} that the
$\rm[4f^{14} 6p]\,\,^2P_{1/2,3/2}$ states will contribute dominantly to the polarizabilities of both the ground and
$\rm[4f^{14} 5d]\,\,^2D_{3/2}$ states. Therefore, knowledge of precise E1 matrix elements between these states are crucial in
estimating polarizabilities of the above two states associated with the clock transition. After discussing accuracies in the static
polarizability results, we move on to presenting the dynamic polarizabilities of the $\rm[4f^{14}6s]\,\, ^2S_{1/2}$ and $\rm[4f^{14}5d]\,\, ^2D_{3/2}$
states in Yb$^+$ and adjudge that these quantities will also have similar accuracies with their corresponding static polarizability
values.

\subsection{Static polarizability}

In Table \ref{polar_ground}, we give contributions from the
individual correlation effects to the static $\alpha_{J_n}^{(i)}$
values of the ground and $ \rm[4f^{14} 5d] ~  ^2 D_{3/2} $ states
in $^{171}$Yb$^+$. The ``Main'' contributions are listed
separately in the table along with the E1 matrix elements and
resonant wavelengths $\lambda_{\rm{res}}$ of different transitions
involving the low-lying $\rm P_{1/2, \, 3/2}$ states to illustrate
their roles explicitly in the accurate evaluation of
$\alpha_{J_n}^{(i)}$. Uncertainties of these matrix elements are given
along with their respective values. The uncertainties in the CCSD
values are determined by estimating contributions from the
neglected leading order triple excitations and due to use of the
the finite size basis functions in the calculation.
We find more than 90\% contribution to the total static $\alpha_{J_n}^{(i)}$ values come from
the $ \rm[4f^{14} 6p] ~  ^2 P_{1/2} $  and $ \rm[4f^{14} 6p] ~  ^2
P_{3/2} $ states in both the cases. Thus, we replace the E1 matrix
elements of the $ \rm[4f^{14} 6s] ~  ^2 S_{1/2} \rightarrow
\rm[4f^{14}6p] ~  ^2 P_{1/2,3/2}$ and $ \rm[4f^{14} 5d] ~  ^2
D_{3/2} \rightarrow \rm[4f^{14}6p] ~  ^2 P_{1/2} $ transitions that are
obtained from the CCSD method by the values extracted from the
combination of the experimental values of the lifetimes and the
corresponding branching ratios from ~\cite{[2],[3]}  of
the $ \rm[4f^{14} 6p] ~ ^2 P_{1/2} $  and $ \rm[4f^{14} 6p] ~  ^2
P_{3/2} $ states. These values are quoted in bold fonts while the
CCSD values are mentioned just below these numbers in Table
\ref{polar_ground}. We, however, could not extract out the E1
matrix element of the $ \rm[4f^{14} 5d] ~  ^2 D_{3/2} \rightarrow
\rm[4f^{14}6p] ~  ^2 P_{3/2} $ transition from any available
experimental data. Nevertheless, we find that the experimental
value of the E1 matrix element, that is 2.97 a.u., corresponding to the $ \rm[4f^{14} 5d] ~  ^2
D_{3/2} \rightarrow \rm[4f^{14}6p] ~  ^2 P_{1/2} $ transition is
very close to our CCSD result (2.95 a.u.). Since the  $\rm[4f^{14} 6p] ~  ^2
P_{1/2} $  and $ \rm[4f^{14} 6p] ~  ^2 P_{3/2} $ states are the
fine structure partners, electron effects will behave similarly in determining these states. Therefore, we
assume that the CCSD value for the E1 matrix element of the $ \rm[4f^{14} 5d] ~  ^2 D_{3/2}
\rightarrow \rm[4f^{14}6p] ~  ^2 P_{3/2} $ transition will have similar accuracy as of the corresponding value of the
$\rm[4f^{14} 5d] ~  ^2 D_{3/2} \rightarrow \rm[4f^{14}6p] ~  ^2P_{1/2} $ transition. Next to the $6P$ states, there lie
the states with $\rm[4f^{13}5d6s]$ configurations. Finding E1 matrix elements for these states are strenuous in the
RCC method, but a CI method can estimate them more reliably. We consider these values
either from the experimental data presented in Ref. \cite{spectrochimica_2010} or are taken from the theoretical calculations
reported in Ref. \cite{[4]} using a CI method. Since uncertainties of these matrix elements obtained from the CI method are not
given and their contributions to the polarizability values are found to be relatively small (see Table \ref{polar_ground}), we have not
accounted for uncertainties from these matrix elements in the present work. In order to infer E1 matrix element of the
$\rm[4f^{13}5d6s] ~^3[3/2]_{1/2} \rightarrow [4f^{14}5d] ~ ^2D_{3/2}$ transition for estimating its contribution in the evaluation
of the polarizability of the $[4f^{14}5d] ~ ^2D_{3/2}$ state, we have used the experimentally measured lifetime
of the $\rm[4f^{13}5d6s] ~^3[3/2]_{1/2}$ state from Ref. \cite{spectrochimica_2010} and branching ratios reported in
Refs. \cite{J_phy_biemont_1998,Hendrik_2012}.

As seen in Table \ref{polar_ground},  the
core correlation has significant contribution to the scalar
$\alpha_{J_n}^{(0)}$ values. This cannot be evaluated
using a sum-over-states approach, so it is imperative to apply a
suitable {\it ab initio} many-body method for its determination. The core correlation contribution $\alpha^{(i,c)}$ to
the ground and $[4f^{14}5d] ~ ^2D_{3/2}$ states of Yb$^+$, which corresponds to dipole polarizability of the Yb $^{2+}$ ion,
has not been evaluated rigorously earlier. To determine this contribution more reliably, we apply three different methods
in the DHF, RPA and CCSD approximations based on the first principle calculations as discussed in Ref. \cite{yashpal}. We find
$\alpha^{(i,c)}=7.45$ a.u., $\alpha^{(i,c)}=6.38$ a.u. and $\alpha^{(i,c)}=7.72$ a.u. from the DHF, RPA and CCSD methods, respectively. In Ref. \cite{MBPT}, the RPA value is given as 6.386 a.u. which agrees well with our RPA value. The DHF value is found to be larger than the RPA value,
the CCSD value is larger than the other two methods. Though RPA is also an all order method, but it takes into account only the
core-polarization correlations while the CCSD method includes pair-correlation effects to all orders along with the core
polarization effects. We, therefore, consider CCSD value as the final result and it is given along with the estimated uncertainty
in Table \ref{polar_ground}.

The ``Tail'' contributions to $\alpha_{J_n}^{(i,v)}$ and
$\alpha_{J_n}^{(i,cv)}$ are also given in Table
\ref{polar_ground}, which are obtained using the DHF method. The
core-valence contributions are found to be negligibly small,
whereas the ``Tail'' contributions to the scalar and tensor
polarizabilities of the $ \rm[4f^{14} 5d] ~ ^2 D_{3/2}$ state are
significant. Accounting all these contributions, we obtain the
final static polarizability of the ground state as 59.3(8) a.u.,
while the static scalar and tensor polarizabilities of the $
\rm[4f^{14} 5d] ~  ^2 D_{3/2}$ state are 107(3) a.u. and $-75(2)$
a.u., respectively. An earlier calculation using the relativistic
many-body perturbation theory (MBPT) presents ground state static
polarizability value as 62.04 a.u.\cite{MBPT}. The discrepancy
arises from the value of E1 matrix elements that are used for the
transitions involving the $ \rm[4f^{14} 6p] ~ ^2 P_{1/2} $ and $
\rm[4f^{14} 6p] ~  ^2 P_{3/2} $ states. On the other hand, Lea
\textit{et. al.} have also estimated this value as 47(9) a.u.
\cite{Polarizability_EFTF} using experimental E1 matrix elements.
In their calculation, only few E1 transitions were included and
the core contribution was completely neglected. Nevertheless, both
the values agree within their respective uncertainties. Accurate
calculations of $\alpha_{J_n}^{(i)}$ for the $\rm [4f^{14}5d] ~
^2D_{3/2}$ state is very challenging as compared to the ground
state. Here, the E1 matrix elements between this state and the
$\rm{[4f^{14}}$ $n\rm p]~  ^2 P_{1/2,3/2}$ and $\rm {[4f^{14}}$ $n
\rm{f]}~ ^2 F_{5/2} $ states, with the principle quantum number
$n$, mainly contribute. We observe from Table \ref{polar_ground}
that contributions from the E1 matrix elements involving the $\rm
{[4f^{14}}$ $n\rm{f]}~ ^2 F_{5/2} $ states converge very slowly as
compared to the $\rm{ [4f^{14}}$ $n \rm{p]} ~ ^2 P_{1/2, 3/2}$
states with higher $n$ value. We, therefore, consider  $n=5,6,7,8$
for the $\rm {[4f^{14}}$ $n\rm{f]}~ ^2 F_{5/2} $ states and
$n=6,7$ for the $\rm{[4f^{14}}$ $n\rm p]~  ^2 P_{1/2, 3/2}$ states
in the evaluation of ``Main'' contribution to
$\alpha^{(i,v)}_{J_n}$. Among these contributions, the E1 matrix
element of the $\rm [4f^{14}5d] ~ ^2D_{3/2}\rightarrow\rm [4f^{14}
6p] ~  ^2 P_{1/2}$ transition dominates. The contribution from the
$\rm [4f^{14}5d] ~ ^2D_{3/2}\rightarrow\rm [4f^{14} 6p] ~ ^2
P_{3/2}$ transition is found to be about ten times smaller than
the above transition. This is owing to the fact that its E1 matrix
element is almost half while its transition energy is twice larger
than the corresponding value of the $\rm [4f^{14}5d] ~ ^2D_{3/2}
\rightarrow\rm [4f^{14}6p] ~  ^2 P_{1/2}$ transition. Noticeably
contributions from the $\rm[4f^{13} 5d 6s]$ configurations are
non-significant as seen in Table \ref{polar_ground}. Unlike for
the ground state, the ``Tail'' contribution to
$\alpha^{(0)}_{J_n}(0)$ is about 4\% of its total value. Thus, the
dominant sources of uncertainty in $\alpha_{J_n}^{(o)}$ comes from
the ``Tail'' contribution. Similar trends are also observed in the
determination of the static $\alpha^{(2)}_{J_n}$ value of the $\rm
[4f^{14}5d] ~ ^2D_{3/2}$ state, which are also quoted in
Table~\ref{polar_ground}. Lea \textit{et. al.} have estimated its
scalar polarizability value as 90(17) a.u.
\cite{Polarizability_EFTF} considering only few  E1 matrix
elements deducing from the experimental lifetime values. The reason for which we believe to obtain more reliable results for $\alpha$ of the ground and $[4f^{14}5d] ~ ^2D_{3/2}$ than the previously estimated values is because of use of more accurate values of the E1 matrix elements of the $\rm [4f^{14}5d] ~ ^2D_{3/2} \rightarrow\rm [4f^{14}6p] ~  ^2 P_{3/2}$, $\rm [4f^{14}5d] ~ ^2D_{3/2} \rightarrow\rm \rm {[4f^{14}}$ $5\rm{f]}~ ^2 F_{5/2}$
and $\rm [4f^{14}5d] ~ ^2D_{3/2} \rightarrow\rm \rm {[4f^{14}}$ $6\rm{f]}~ ^2 F_{5/2}$ transitions, and core contribution that are evaluated using our RCC method. An experimental value of the tensor polarizability of the
$\rm[4f^{14} 5d] ~  ^2 D_{3/2}$ state has been reported as
$-82(13)$ a.u. \cite{experimental_polarization}. From the
comparison, we find our estimated value agrees with the
experimental result indicating that we are able to produce
polarizabilities in the investigates states accurately.

Using the static polarizability values as mentioned above, we estimate the BBR shifts in the ground and $\rm[4f^{14}5d] ~ ^2D_{3/2}$
states of Yb$^{+}$ as $-0.52$ Hz and $-0.92$ Hz, respectively. Further, we evaluate the dynamic corrections to these quantities
approximately, using Eq. (\ref{dync}), as $-0.0004$ Hz and $-0.04$ Hz in the ground and $[4f^{14}5d] ~ ^2D_{3/2}$
states respectively. From this analysis, we obtain the net differential scalar polarizability value of the $\rm[4f^{14}6s] ~
 ^2S_{1/2} \rightarrow [4f^{14}5d] ~ ^2D_{3/2}$ transition as $-7.8$(5)$\,\times\, 10^{-40}\,\rm Jm^2 V^{-2}$ against the
available experimental value of $-6.9$(1.4)$\,\times\, 10^{-40}$\, $\rm Jm^2V^{-2}$ \cite{experimental_polarization}. This corresponds
to the differential BBR shift for the $\rm[4f^{14}6s] ~^2S_{1/2} \rightarrow [4f^{14}5d] ~ ^2D_{3/2}$ clock transition as -0.44(3) Hz.
This value comes out to be larger than the measured value $-0.37(5)$ Hz by Tamm  et al  ~\cite{tamm_eftf}, which is being currently considered
for estimating uncertainty due to the BBR shift in the above clock transition frequency. There are also two different values
were predicted as $-0.254$ Hz and $-0.261(8)$ Hz by Lea et al from the {\it ab initio} calculations and combining with both the
experimental and theoretical data, respectively \cite{Polarizability_EFTF}.

\subsection{Magic wavelengths for linearly and circularly polarized light}

\begin{figure*}[t]
\begin{center}
\includegraphics[width=16.0cm,height=6.0cm]{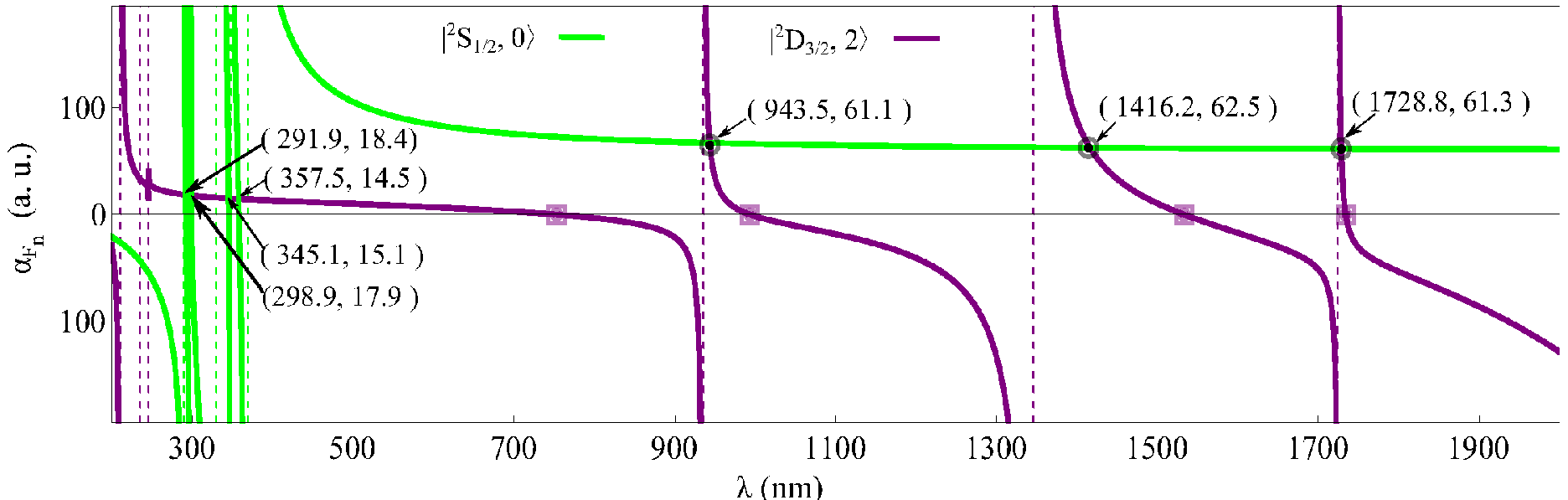}
\end{center}
\caption{Dynamic scalar polarizabilities of the $\rm [4f^{14}6s] ~
^2S_{1/2}$ and $\rm [4f^{14}6s]\,^2D_{3/2}$ states when vector and
tensor polarizability components are zero. The
$\lambda_{\rm{magic}}$ and  $\lambda_{\rm{T}}$ wavelengths {\bf for comprehensive trapping} are
pointed out in dots and encircled by circles and squares,
respectively, are independent of the hyperfine levels and magnetic
sub-levels of the states. In particular to the
$\lambda_{\rm{magic}}$, their corresponding $\alpha_{F}$ values
are indicated within brackets.The vertical dashed lines signify the
resonances from the ground and
$\rm[4f^{14}5d] ~^2D_{3/2}$ states in green and purple
colors, respectively.  \label{magic_insensitive}}
\end{figure*}

In order to find out $\lambda_{\rm{magic}}$ among different
magnetic sublevels $M_{F}$ of the ${\rm[4f^{14}6s] ~
^2S_{1/2}}(F=0) \rightarrow {\rm [4f^{14}5d] ~ ^2D_{3/2}}(F=2)$ clock
transition we estimate dynamic $\alpha_{F_n}$ values of the associated states at different
wavelengths ranging from 200-2000 nm. First the atomic polarizabilities of the
$\rm [4f^{14}6s] \,\, ^2S_{1/2}$ and $\rm[4f^{14}5d]\,\, ^2D_{3/2}$ states are determined using
the relations given in Eqs.~(\ref{scalar}-\ref{tensor}), then total polarizability value is obtained from Eq.~(\ref{total}).
In Fig.\ref{linear1} we plot our estimated $\alpha_{F_n}$ values of
both the states as function of the wavelength. To obtain the
$\rm{\lambda_{magic}}$ values for the $\rm|4f^{14}6s\,\,
^2S_{1/2},\,\it{F}=\rm{0}\rangle \rightarrow |\rm 4f^{14}5d\,\,
^2D_{3/2},\,\it{F}=\rm{2} \rangle$ transition, we have considered
all possible combinations of $\it {M_F}$ sublevels and also
different light polarizations. In Fig.~\ref{linear1}(a), we plot
the dynamic polarizabilities for the states associated with the $|\rm^2S_{1/2},\it{F}=\rm0\rangle \rightarrow |\rm
^2D_{3/2},\,\it{F}=\rm2,\,\it{M_F}=\rm0, \pm\,1, \pm\,2\rangle$
 transitions against the wavelength corresponding to a linearly polarized light.
Similar plots are shown in Fig. \ref{linear1}(b) for the
$\rm|^2S_{1/2},\it{F}=\rm0\rangle$ and
$\rm|^2D_{3/2},\,\it{F}=\rm2,\,\it{M_F}=\rm-2,\,-1,\,0,\,1,\,2\rangle$
states assuming a circular polarized light. {\bf The wavelengths at which intersections of the
polarizability curves take place are recognized as $\rm \lambda_{\rm{magic}}$.} In Table
~\ref{magic} we list the $\lambda_{\rm{magic}}$ values for the
$\rm|4f^{14}6s\,\, ^2S_{1/2},\,\it{F}=\rm{0}\rangle \rightarrow
|\rm 4f^{14}5d\,\, ^2D_{3/2},\,\it{F}=\rm{2} \rangle$ transition
in the presence of a linearly polarized light. It can be noticed
that these $\lambda_{\rm{magic}}$ values support a red-detuned
trap (depicted by the positive values of dipole polarizabilities
at these wavelengths) except at 245.4 nm. Magic wavelengths at the
infrared region are considered to have more experimental
significance due to availability of high power fiber or solid
state lasers in comparison to ultraviolet wavelengths produced by
the sum or difference frequency generation techniques. Similarly,
the $\lambda_{\rm{magic}}$ for the $\rm|4f^{14}6s\,\,
^2S_{1/2},\,\it{F}=\rm{0}\rangle \rightarrow |\rm 4f^{14}5d\,\,
^2D_{3/2},\,\it{F}=\rm{2} \rangle$ transition in the presence of a
right-circularly polarized light are also given in Table
~\ref{magic}. In the present work, we determine
$\lambda_{\rm{magic}}$ for right circularly polarization using $A
= 1$ considering all possible positive and negative $M_F$
sublevels of the states participating in the transition. Note that
$\lambda_{\rm{magic}}$ for the left circularly polarized light of
a transition with a given $M_F$ are equal to right circularly
polarized light with opposite sign of $M_F$
%%%%%%%%%%%%%%%%%%%%%%%%%%%%%%%%%%%%%%%%%%%%%%%%%%%%%%%%%%%%%%%%%%%%%%%%%%%%%%%%%%%%%%%%%%%%%%%%
\begin{table*}
\centering{\caption{The $\lambda_{\rm{T}}$ values in nm  for all
possible $M_{F}$ components of the considered states in
$^{171}$Yb$^+$ using linearly and right-circularly
polarized light. {\bf The term "Both" for ${\rm{[4f^{14}6s] ~ ^2S_{1/2}}}(F=0)$ state means that the results are same for linear and circular polarization.} \label{tune} }
% \begin{ruledtabular}
\begin{tabular}{lrrlrrlrr}
\hline \hline
   \multicolumn{3}{c}{${\rm{[4f^{14}6s] ~ ^2S_{1/2}}}(F=0)$ } &  \multicolumn{3}{c}{${\rm{[4f^{14}5d] ~ ^2D_{3/2}}}(F=2)$ }\\
   \hline
Polarization   & $M_{F}$  &  $\lambda_T$\,\,\,\,\,\, & Polarization   & $M_{F}$  &  $\lambda_T$\,\,\,\,\,\, & Polarization   & $M_{F}$  &  $\lambda_T$\,\,\,\,\,\,\\
 \hline
 Both &  0\,\,\,\,  &   357.9(1) & Linear       & 0\,\,\,\,           & 1725.4(1)  & Circular    & -2\,\,\,\,          & 1726.9(1)  \\
 &  \,\,\,\,  &   345.2(1) &             &             & 1370.2(3)                  &               &             & 1398.1(2) \\
 &            &   299.1(1)          &             &            & 1026.9(8)                  &                &             & 961.7(1)           \\
 &  \,\,\,\,  &  292.2(1) &             &             &  741.2(5)                    &                 &             & 283.3(4)   \\
 &  \,\,\,\,  &           &             &  $\pm1$\,\,\,\,     & 1727.8(1)           &             & -1\,\,\,\,          & 1731.3(1)  \\
 &            &            &             &                      & 1438.6(3)           &            &             &  1478.1(1)  \\
 &            &            &              &                      & 1021(1)            &             &             & 961.6(1)         \\
 &             &            &               &                    & 774(4)             &             &                    &  468.2(7)  \\
 &            &             &               & $\pm2$\,\,\,\,      & 1712.4(3)         &             & 0\,\,\,\,           & 1826(2)   \\
  &           &             &               &                     & 966(4)            &              &                      & 1688.3(3)  \\
  &           &             &                &                    &                     &              &                    & 1000(2) \\
   &          &              &               &                    &                    &              &                     & 861(3)  \\
   &          &              &               &                    &                    &               &  1\,\,\,\,           & 1719.4(1)   \\
   &           &             &                &                    &                   &                  &                     & 1233(1)  \\
   &            &              &               &                     &                 &                   & 2\,\,\,\,           & -  \\
\hline \hline
\end{tabular}}
% \end{ruledtabular}
\end{table*}
%%%%%%%%%%%%%%%%%%%%%%%%%%%%%%%%%%%%%%%%%%%%%%%%%%%%%%%%%%%%%%%%%%%%%%%%%%%%%%%%%%%%%%%%%%%%%%%%%%

\subsection{Magic wavelengths for comprehensive trapping}

In Fig.~\ref{magic_insensitive}, we show $\lambda_{\rm magic}$
values that are independent of the hyperfine sub-levels and their $
{M_F}$ sublevels of the states involved in the clock transition.
These $\lambda_{\rm{magic}}$ values will produce null differential
Stark shifts exclusively among any hyperfine level of the ground
and excited state. We have identified at least  seven magic wavelengths for
the hyperfine level independent trapping, out of which four are in the
ultra violet and two in the infra red region. The
$\lambda_{\rm{magic}}$ values along with the polarizability values
at these wavelengths are marked with arrows in Fig.
~\ref{magic_insensitive}. In order to distinguish the $M_F $
independent magic wavelength, the $\lambda_{\rm{magic}}$ presented
in Table~\ref{magic} for linearly and circularly polarized light
can be compared to the values marked in
Fig.~\ref{magic_insensitive}. The trapping scheme as proposed in
Ref.~\cite{sukhjit} and using $\lambda_{\rm{magic}}$ indicated in
Fig.~\ref{magic_insensitive}, one can optically trap $\rm{Yb}^+$
with zero light shift in the $\rm [4f^{14}6s]\,\,^2S_{1/2}
\rightarrow \rm[4f^{14}5d]\,\,^2D_{3/2}$ transition for any given combinations
of the $F$ and $M_F$ levels that are experimentally viable.

%%%%%%%%%%%%%%%%%%%%%%%%%%%%%%%%%%%%%%%%%%%%%%%%%%%%%%%%%%%%%%%%%%%%%%%%%%%%%%%%%%%%%%%%%%%%%%%%%%%%%%%%%%%%%%%%%%
%\begin{table}
%\centering{\caption{ Magic wavelengths for hyperfine level independent
%trapping of $|\rm 4f^{14}6s\,\,^2S_{1/2}\rangle \rightarrow |\rm
%4f^{14}5d\,\,^2D_{3/2}\rangle$ transition. \label{magic_scalar}}
%% \begin{ruledtabular}
%\begin{tabular}{lrrrrr}
%\hline \hline
%      $\lambda_{\it{magic}}$\,\,\,\  &$\alpha$\,\,\, \,\,\,         \\
%      \,\,\,\ nm            & a.u. \,\,\,\                          \\
% \hline
%                            &                             \\
%      1728.7(1) ~ \,                   & 61.4(1)                 \\
%     1414.4(3) ~ \,                  & 62.5(2)                        \\
%   357.5(7) ~ \,                   &  14.8(9)           \\
%   345.1(4) ~ \,                   &  15.3(8)          \\
%   298.8(1) ~ \,                   &  17.9(1.3)          \\
%    291.9(1) ~ \,                   &  18.6(9)          \\
%\hline \hline
%\end{tabular}}
%% \end{ruledtabular}
%\end{table}
%%%%%%%%%%%%%%%%%%%%%%%%%%%%%%%%%%%%%%%%%%%%%%%%%%%%%%%%%%%%%%%%%%%%%%%%%%%%%%%%%%%%%%%%%%%%%%%%%%%%%%%%%%%%%%%%%%%%%

\subsection{Tune-out wavelengths}

We have identified  
polarization dependent 
tune-out wavelengths at
which the dynamic polarizabilities reduces to zero individually with different $M_F$ sub-levels in both
$F=0$ state of $\rm[4f^{14}6s]\,\, ^2S_{1/2}$ and $F=2$ state of $\rm[4f^{14}5d]\,\, ^2D_{3/2}$. These $\lambda_{\rm T}$ values
are listed in Table~\ref{tune} together with the $M_F$ values against wavelengths of light with linear and circular
polarizations. We have also identified four tune-out wavelengths for the $\rm[4f^{14} 5d]\,\,^2D_{3/2}$
state at 1734.2(2), 1533.2(1), 993.5(5), and 753(3) nm, which could be useful for carrying out measurements irrespective of choice
for hyperfine sub-levels.

\section{Conclusion}

In summary, we have conducted a systematic study of the static polarizabilities for the ground  $\rm[4f^{14}6s]~^2S_{1/2}$ and metastable
$\rm [4f^{14}5d]~^2D_{3/2}$ states of Yb$^+$. The polarizability values are further
used to estimate the BBR shift for the  $\rm[4f^{14}6s]~^2S_{1/2}\rightarrow [4f^{14}5d]~^2D_{3/2}$
transition. The values of the ac polarizabilities are also determined for wavelengths ranging from the ultraviolet through infrared spectral
regions, and are used to find out the magic wavelengths for optical trapping of the above mentioned clock transition. These magic
wavelengths are given for both the linearly and circularly polarized light and also for a comprehensive trapping scheme which is
independent of the hyperfine sub-levels of the states involved in the clock transition. We were also able to identify a number of tune-out
wavelengths for the above states. Our results will be of interest for carrying out precision measurements in the
$\rm[4f^{14}6s]~^2S_{1/2} \rightarrow [4f^{14}5d]~^2D_{3/2}$ transition of Yb$^+$ for attaining high accuracy clock frequency and
investigating many fundamental physics.

\section*{Acknowledgements}
S.D. acknowledges CSIR-National Physical Laboratory, Department of
Science and Technology (grant no. SB/S2/LOP/033/2013) and Board of
Research in Nuclear Sciences (grant no. 34/14/ 19/2014-BRNS/0309)
for supporting this work. The work of B.A. is supported by
Department of Science and Technology, India. B.K.S. acknowledges
support under the PRL-TDP program and use of Vikram-100 HPC
Cluster at Physical Research Laboratory, Ahmedabad.


\section*{References}

\begin{thebibliography}{}
\bibitem{roberts} M. Roberts  {\it et. al.}, Phys. Rev. A {\bf 62}, 020501(R) (2000).
\bibitem{tamm}  Chr. Tamm, S. Weyers, B. Lipphardt and E. Peik, Phys. Rev. A {\bf 80}, 043403 (2009).
\bibitem{imai} Y. Imai, K. Sugiyama, T. Nishi, S. Higashitani, T. Momiyama and M. Kitano, Poster No. B3-PWe21  {\it The $12^{th}$ Asia Pacific Physics Conference}, 14-19 July 2013.
\bibitem{atish} A. Rastogi {\it et .al}., Mapan {\bf 30}, 169 (2015)
\bibitem{huntemann} N. Huntemann, C. Sanner, B. Lipphardt, Chr. Tamm, and E. Peik, Phys. Rev. Lett. {\bf 116}, 063001 (2016).
\bibitem{bijaya} B. K. Sahoo and B. P. Das, Phy. Rev. A {\bf 84}, 010502(R) (2011).
\bibitem{dzuba1}
V. A. Dzuba, V. V. Flambaum, and M. V. Marchenko, Phys. Rev. A {\bf 68}, 022506 (2003).
\bibitem{dzuba2}
V. A. Dzuba and V. V. Flambaum, Phys. Rev. A {\bf 77}, 012515 (2008).
\bibitem{godun} R. M. Godun {\it et. al}., Phys. Rev. Lett. {\bf 113}, 210801 (2014).
\bibitem{dzubanat} {V. A. Dzuba, V. V. Flambaum, M. S. Safronova, S. G. Porsev, T. Pruttivarasin, M. A. Hohensee, and H. Haffner, Nat. Phys. {\bf12}, 465 (2016).}
\bibitem{yu} N. Yu and L. Maleki, Phys. Rev. A {\bf 61}, 022507 (2000).
\bibitem{nandy}{ D. K. Nandy and B. K. Sahoo, Phys. Rev. A {\bf 90}, 050503(R) (2014)}.
\bibitem{neha_2016}{N. Batra, B. K. Sahoo and S. De., Chin. Phys. B. {\bf25}, 113703 (2016).}
\bibitem{Polarizability_EFTF}{S. N. Lea, S. A. webster, G. P.  Barwood,  {\it Proceeding of the $20^{th}$ European Frequency and Time Forum (EFTF)}, edited by F. Riehle (PTB, Braunschweig, Germany, 2006), pp. 302-307.}
\bibitem{tamm_eftf}{Chr. Tamm {\it et. al.}, {\it Proceeding $20^{th}$ European Frequency and Time Forum (EFTF)}, edited by F.Riehle(PTB, Braunschweig, Germany, 2006).}
\bibitem{experimental_polarization}{T. Schneider, E. Peik and C. Tamm, Phys. Rev. Lett. {\bf94}, 230801 (2005).}
\bibitem{PLLiu_PRL_2015}{Pei-Liang Liu, {\it et. al}. Phys. Rev. Lett. {\bf114}, 223001 (2015).}
\bibitem{ChSchnieder_Nature_2010}{Ch. Schnieder, M. Enderleiin, and T. Schaetz, Nature. Photonics {\bf4}, 772 (2010).}
\bibitem{MEnderlein_PRL_2012}{Martin Enderlein, Thomas Huber, Christian Schneider, and Tobias Schaetz, Phys. Rev. Lett. {\bf109}, 233004 (2012).}
\bibitem{SDiang_PRL_2014}{Shiqian Ding, Huanqian Loh, Roland Hablutzel, Meng Gao, Gleb Maslennikov, and Dzmitry Matsukevich, Phys. Rev. Lett. {\bf113}, 073002 (2014).}
\bibitem{zhang_2017} {J. Zhang, {\it et. al}, Nature {\bf 543}, 217 (2017).}
\bibitem{katori123} {Hidetoshi Katori, Koji Hashiguchi, E. Yu. Ilinova and V. D. Ovsiannikov, Phys. Rev. Lett. {\bf 103}, 153004 (2009)}.
\bibitem{andrei}  {Andrei Derevianko and Hidetoshi Katori, Rev. Mod. Physics {\bf83}, 331 (2011)}.
\bibitem{barber} Z. W. Barber, C. W. Hoyt, C. W. Oates, L. Hollberg, A. V. Taichenachev, and V. I. Yudin, Phys. Rev. Lett. {\bf96}, 083002 (2006).
\bibitem{bindiya1234}  Bindiya Arora, M. S. Safronova, and Charles W. Clark, Phys. Rev. A {\bf76}, 052509 (2007).
\bibitem{JKaur_PRA_2015}{Jasmeet Kaur, Sukhjit Singh, Bindiya Arora, and B. K. Sahoo, Phys. Rev. A {\bf92}, 031402 (2015).}
\bibitem{tuneout}{Bindiya Arora, M. S. Safronova, and Charles W. Clark, Phys. Rev. A {\bf84}, 043401 (2010).}
\bibitem{sherman}
J. A. Sherman, T. W. Koerber, A. Markhotok, W. Nagourney, and E. N. Fortson, Phys. Rev. Lett. {\bf 94}, 243001 (2005).
\bibitem{bkslight}
B. K. Sahoo, L. W. Wansbeek, K. Jungmann, and R. G. E. Timmermans, Phys. Rev. A {\bf 79}, 052512 (2009).
\bibitem{beloy_2009}{K. Beloy, {\it Theory of  the ac stark effect on the atomic hyperfine structure and applications to microwave atomic clocks} (2009), PhD thesis submitted to University of Nevada, Reno, USA.}
\bibitem{Dzuba_2010} {V. A. Dzuba, V. V. Flambaum, K. Beloy, and A. Derevianko, Phys. Rev. A {\bf82}, 062513 (2010).}
\bibitem{porsev_2006}{S. G. Porsev and A. Derevianko, Phys. Rev. A {\bf74}, 020502(R) (2006).}
\bibitem{Manakov_1986} {N. L. Manakov, V. D. Ovsiannikov, and L. P. Rapport, Phys. Rep. {\bf141}, 319 (1986).}
\bibitem{b.arora.nandy}{Bindiya Arora, D. K. Nandy and B. K. Sahoo, Phys. Rev. A {\bf85}, 012506 (2012).}
\bibitem{MBPT}{U. I. Safronova and M. S. Safronova, Phys. Rev. A {\bf79}, 022512 (2009).}
\bibitem{[2]}{S. Olmschenk, K. C. Younge, D. L. Moehring, D. N. Matsukevich, P.Maunza and C.Monroe, Phys. Rev. A {\bf80}, 022502 (2009)}
\bibitem{[3]}{E. H. Pinnington, G. Rieger, and J. A. Kernahan, Phys. Rev. A {\bf3}, 562421 (1997)}
\bibitem{[4]}{S. G. Porsev, M. S. Safronova and M. G. Kozlov, Phys. Rev. A {\bf 86}, 022504 (2012)}.
\bibitem{spectrochimica_2010}{D. Kedzierski, J. Kusz, J. Muzolf, Spectrochimica Acta Part B. {\bf65}, 248-252 (2010).}
\bibitem{J_phy_biemont_1998}{E. Biemont, J. F. Dutrieux, I. Martin, and P. Quinet, J. Phys. B {\bf31}, 3321 (1998).}
\bibitem{Nist_data}{NIST atomic spectra Database, $\rm http://www.nist.gov/pml/data/asd.cfm$}
\bibitem{yashpal} Y. Singh, B. K. Sahoo, and B. P. Das, Phys. Rev. A {\bf88}, 062504 (2013).
\bibitem{sukhjit} S. Singh, B. K. Sahoo, and B. Arora, Phys. Rev. A {\bf 93}, 063422 (2016).
\bibitem{Hendrik_2012}{Hendrik M. Meyer, Matthias Steiner, Lothar Ratschbacher, Christoph Zipkes, and Michael Kohl, Phys. Rev. A {\bf85}, 012502 (2012) .}




%\bibitem{dow} J. M. Dow, R. E. Neilan, and C. Rizos, J. Geod. {\bf 83}, 191 (2009).
%\bibitem{white} J. White and R. Beard, {\it  Proceedings of the 33rd Annual Precise Time and Time Interval (PTTI) Systems and Applications Meeting}, Long Beach, California, USA (U.S. Naval Observatory, Washington, D.C. 2002).
%\bibitem{ramsey} N. F. Ramsey, {\it Applications of Atomic Clocks} (Laser physics at the limits, Springer publication, 2002).
%\bibitem{normile} D. Normile and D. Clery, Science {\bf 333}, 1820 (2011).
%\bibitem{uzan} J. P. Uzan, Rev. Mod. Phys. {\bf 75}, 403 (2003).
%
%
%\bibitem{olmschenk} S. Olmschenk {\it et. al}., Phys. Rev. A {\bf 76}, 052314 (2007).
%\bibitem{bloom} B. J. Bloom {\it et. al}., Nature {\bf 506}, 71 (2014).
%
%\bibitem{Lisdat} C. Lisdat, {\it et. al.}, arXiv:1511.07735 (2015).
%\bibitem{Jean} Jean-Daniel Deschnes {\it et. al}., Phys. Rev. X {\bf 6}, 021016 (2016).
%
%\bibitem{masao123} Ichiro Ushijima, Masao Takamoto, Manoj Das, Takuya Ohkubo and Hidetoshi Katori, Nature. Photonics {\bf 9}, 185179 (2015).
%\bibitem{katori321}  Hidetoshi Katori, Nature Photonics {\bf 5}, 203170 (2011).
%
%\bibitem{tomoya} Tomoya Akatsuka, Masao Takamoto, and Hidetoshi Katori, Phys. Rev. A {\bf81}, 023402 (2010).
%\bibitem{vitali} Vitali D. Ovsiannikov, Andrei Derevianko, and Kurt Gibble, Phys. Rev. Lett. {\bf107}, 093003 (2011).
%\bibitem{mejri} L. Yi, S. Mejri, J. J. McFerran, Y. Le Coq, and S. Bize, Phys. Rev. Lett. {\bf106}, 073005 (2011).
%\bibitem{masao}  Masao Takamoto, Feng-Lei Hong, Ryoichi Higashi1 and Hidetoshi Katori, Nature {\bf435}, 321 (2005).
%\bibitem{andrei123}  Andrei Derevianko, {\it et. al}., Phys. Rev. Lett. {\bf105}, 033002 (2010).
%
%\bibitem{ref2} D. Mukherjee, B. K. Sahoo, H. S. Natraj  and B. P. Das, J. Phys. Chem. A {\bf113}, 12549 (2009).
%\bibitem{bijaya3} B. K. Sahoo, S. Majumder, R. K. Chaudhuri, B. P. Das and D. Mukhrejee, J. Phys. B {\bf37}, 3409 (2004).
%\bibitem{[1]}{S. Olmschenk, K. C. Younge, D. L. Moehring, D. N. Matsukevich, P. Maunza and C. Monroe, Phys. Rev. A {\bf76}, 052314 (2007)}
%
%
%\bibitem{b.arora.safronova}{Bindiya Arora, M. S. Safronova and Charles W. Clark, Phys. Rev. A {\bf84}, 043401 (2011).}
%
%\bibitem{bindiya456}{Bindiya Arora  \,and  B. K. Sahoo, Phys. Rev. A {\bf86}, 033416 (2012).}
%
%
%%
%
%%
%%
%%\bibitem{[5]}{N.L.Manakov, V.D.Ovsiqnnikov and L.P.Rapoport, Phys.Rev.rep.141319(1986)}
%%
%%\bibitem{[6]}{K.D.Bonin and V.V.Kresin, Electric dipole polarizability of atoms, molecules and clusters(world scientific,Singapore,1997)}
%%
%\bibitem{dfccsd}{Bindiya Arora \,and \,B. K. Sahoo, Phys. Rev. A {\bf86}, 033416 (2012).}
%%
%%\bibitem{[8]}{B.K.Sahoo, B.P.Das ( Theoretical studies of the long lifetime of the 6d$^2D_{3/2,5/2}$ states in Fr: Implication for parity nonconservation measurement)}
%

%


%

%


%

%
%\bibitem{CTSchmiegelow_PRL_2016}{C. T. Schmiegelow, {\it et. al}. Phys. Rev. Lett. {\bf116}, 033002 (2016).}
%

%

%

%

%\bibitem{quantum phase}{New J. Physics 10, 045017 (2008)}
%
%\bibitem{quantum information}{Phys. Rev. Lett. 87, 257904 (2001)}
%
%\bibitem{quant inform}{Nature 417, 709 (2002)}
 %
 %\bibitem{quantum magnetism}{New J. Phys 10, 045017 (2008)}
 %
% \bibitem{quant_magnet}{New J. Phys 13, 075012 (2011)}
 %
 %\bibitem{Single ion}{Phys. Rev. Lett. 116, 063001 (2016)}
 %
% \bibitem{paul traps}{New J. Phys, 13, 075012 (2011)}
 %
 %\bibitem{space charge pptential}{"optical pumping of stored atomic ions", D.J.Wineland, W.M.Itano, J.C.Bergquist, J.J.Bollinger and J.D.Prestage, 1985,  Ann.Phys.Fr.10(1985)737-748}
%
%\bibitem{quadrupole moment yb ion}{T. Schneider, E. Peik, and Chr. Tamm., Phys. Rev. Lett.94, 230801 (2005).}
%
%\bibitem{Periodicity of ions}{Phys.Rev.Lett. 116, 033002 (2016)}
%


\end{thebibliography}
\end{document}